\documentclass[aip,apl,reprint,fleqn]{revtex4-1}

\usepackage{graphicx}
\usepackage{amsmath}
\usepackage{epstopdf}

\begin{document}

\title{Probing the Free-carrier Absorption in Multi-Layer Black Phosphorus}

\author{Y. Aytac}
\email[Electronic mail: ]{yaytac@umd.edu}
\affiliation{Institute for Research in Electronics \& Applied Physics, University of Maryland, College Park, MD 20742, USA}
\author{M. Mittendoff}
\affiliation{Fakult\"{a}t f\"{u}r Physik, Universit\"{a}t Duisburg-Essen, Lotharstr. 1, 47057 Duisburg, Germany}
\author{T. E. Murphy}
\affiliation{Institute for Research in Electronics \& Applied Physics, University of Maryland, College Park, MD 20742, USA}

\date{\today}

\begin{abstract}
We study the carrier relaxation dynamics in thin black phosphorus (bP) using time-resolved differential transmission measurements.  The inter-band and intra-band transitions, relaxation, and carrier recombination lifetimes are revealed by tuning the mid-infrared probe wavelength above and below the bandgap of black phosphorus. When the probe energy exceeds the bandgap, Pauli blocked inter-band transitions are observed.  The differential transmission signal changes sign from positive to negative when the probe energy is below the bandgap, due to the absence of inter-band transitions and enhancement in the free-carrier absorption (FCA). The minority carrier lifetime and radiative recombination coefficient are estimated 1.3 ns, and 5.9$\rm \times 10^{-10}$  $\rm cm^{3}/s$, respectively. The overall recombination lifetime of bP is limited by radiative recombination for excess carrier densities larger than 5$\rm \times 10^{19}$ $\rm cm^{-3}$.
\end{abstract}

\maketitle

Black phosphorus (bP) is an atomically thin allotrope of phosphorus with a direct bandgap that varies from 2 eV (0.62 $\rm \mu$m) in monolayers to 0.3 eV (4.0 $\mu$m) for multi-layer ($>5$) films\cite{Li.2017,Tran.2014}.  Black phosphorus is a versatile material for many opto-electronic applications due to its tunable bandgap, high carrier mobility, and optical and electrical anisotropy \cite{Qiao.2014,Xia.2014,Engel.2014,Youngblood.2015,wang2016large}.  While bP is widely regarded as a promising new narrow-bandgap material with numerous applications in the mid-infrared (MIR) regime, prior pump-probe measurements of bP have mainly utilized near-infrared (NIR) photon energies that are far above the bandgap \cite{youngblood2016layer,suess2015carrier,ge2015dynamical,iyer2017mid,suess2016mid}. These early measurements reveal that the carrier dynamics in multi-layer bP are governed by an interplay between inter-band and intra-band processes with comparable strength and differing time-scales \cite{suess2015carrier,wang2016ultrafast}.  One notable recent study employed transient reflection measurements at wavelengths out to 4.7 $\mu$m, which is close to the predicted bandgap\cite{iyer2017mid}.

In this $Letter$, the MIR probe pulses are specifically tuned from 2.4 $\mu$m to 5.5 $\mu$m, and measured in transmission, which provides a direct and unambiguous assessment of the carrier dynamics in this technologically relevant spectral regime. When the material is probed with photon energies larger than the multi-layer bandgap of 300 meV, fast pump-induced saturable absorption is observed that is caused by transient Pauli blocking. On the other hand, when the probe energy is smaller than the multi-layer bP bandgap, we observe a negative pump-induced differential-transmission signal that is attributed entirely to free carrier absorption (FCA). In this case, the carrier recombination lifetime of multi-layer bP can be directly observed without the contribution of the inter-band transitions.  By examining the fluence dependence and non-exponential decay, we infer the role of radiative and non-radiative carrier recombination processes.
%To fully understand the influence of structural anisotropy on carrier lifetime, co- and cross-polarized time-resolved differential-transmission measurements are performed at armchair and zig-zag directions of bP at 5.5 $\mu$m probe wavelength. A longer carrier recombination lifetime is observed at the co-polarized pump-probe configuration along with the arm-chair direction.

\begin{figure}[htb]
 \includegraphics[width=3in, angle=0]{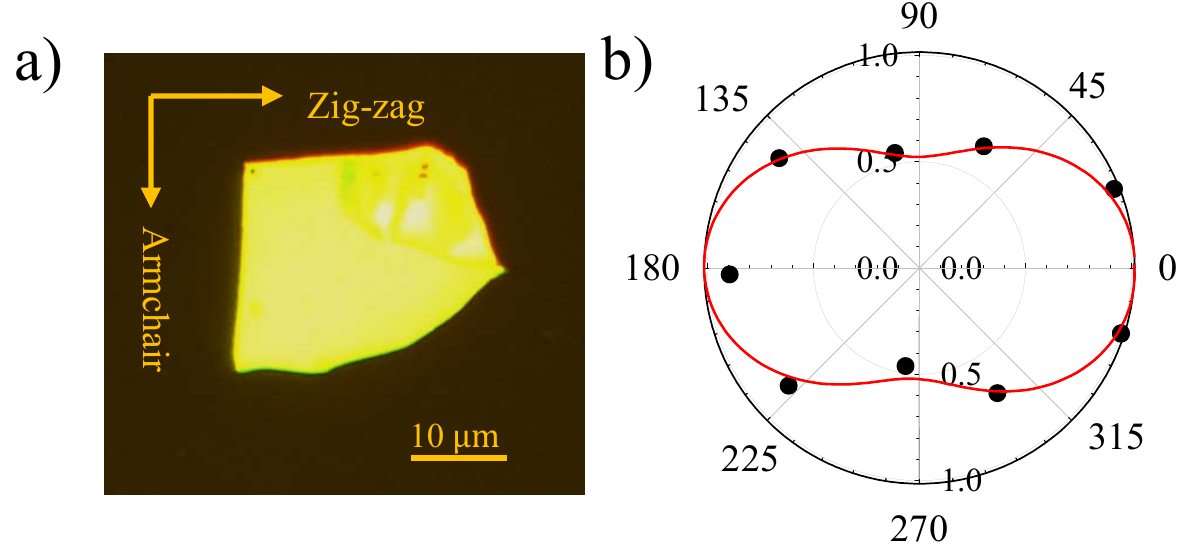}
 \caption{\label{Figure_1} (a) The microscopic image of the bP flake used for the measurements. (b) Polarization-dependent linear transmission measurements at 1.55 $\mu$m wavelength. The data is normalized to the strongest transmission at  0-degree which is parallel to the zig-zag direction as indicated in (b). }
\end{figure}

\begin{figure*}[htb]
 \includegraphics[width=6in, angle=0]{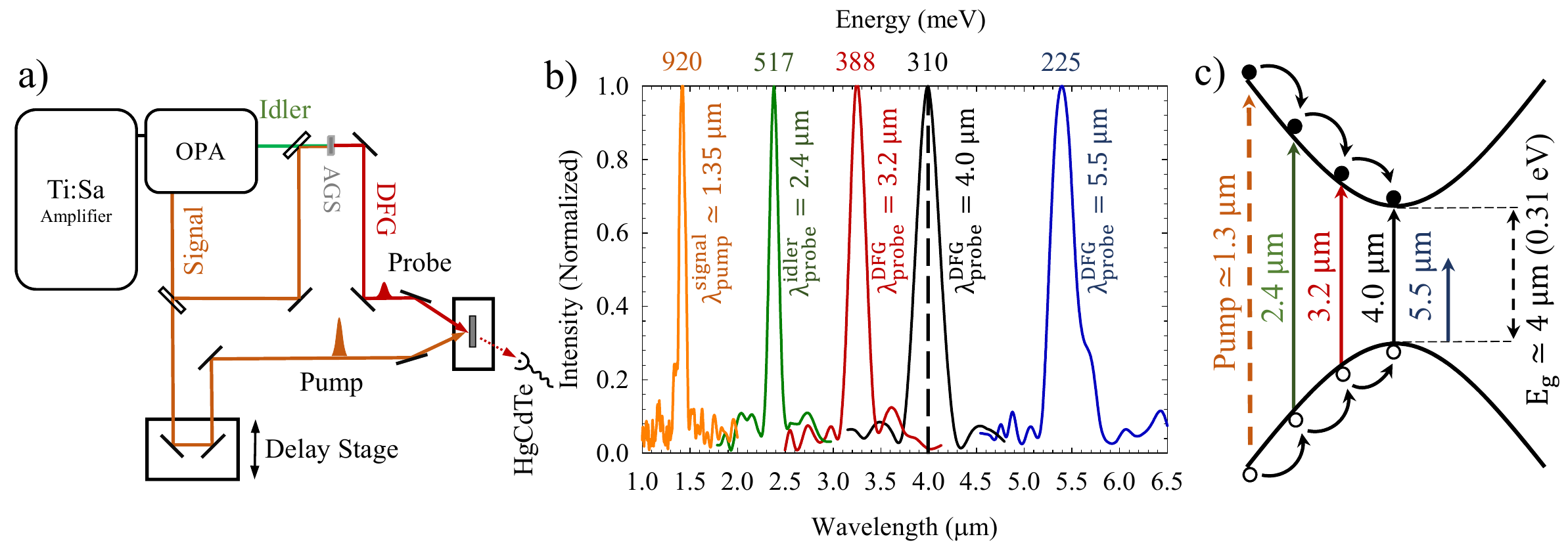}
 \caption{\label{Figure_2} (a) A schematic diagram of the synchronized, two-color femto-second pump-probe system. (b) Pump and probe spectra as obtained from Fourier Transform-IR measurements. The dashed line indicates bandgap of multi-layer black phosphorus at room temperature. (c) A schematic diagram of the photon energies in relation to the band structure represents the sensitivity of probe pulse with corresponding absorptions.}
\end{figure*}

The bP flakes were produced by mechanical exfoliation from bulk, and transferred to a 400 $\rm \mu$m thick CaF$_2$ substrate, which is transparent in the mid-IR range.  Immediately following exfoliation, the samples were encapsulated using an approximately 200 nm thick layer of PMMA produced by spin coating technique. The purpose of the PMMA layer is to prevent oxidation of the bP flakes in ambient conditions during the measurements \cite{Engel.2014,bhaskar2016radiatively}. Fig.~\ref{Figure_1}(a) shows a representative bP flake, with dimensions of 30$\times$40 $\mu$m$^2$. To determine the orientation of the flake, we performed polarization dependent transmission measurements at a wavelength of 1.55 $\mu$m. Figure \ref{Figure_1}(b) shows a polar plot of the measured normalized transmission, as a function of the direction of linear polarization. According to these measurements, the zig-zag direction of the flake is horizontally aligned. The orientation result was further verified by observing the polarization dependence of third-harmonic generation \cite{youngblood2016layer,autere2017rapid} from the flake. The thickness of the bP flake was estimated to be 150 nm, based upon the measured optical transmission of a focused laser at 1.55 $\mu$m \cite{suess2016mid,wang2016ultrafast,castellanos2014isolation,li2015polarization,zhang2017infrared}.

The non-equilibrium carrier dynamics were investigated by using time-resolved differential transmission measurements, in which a strong pump pulse produces a transient change in the optical transmission of a weaker, mid-IR probe pulse.  The pump pulses were generated using a 1 kHz regeneratively amplified Ti-sapphire laser and an optical parametric amplifier, which produces tunable complementary signal and idler wavelengths. A portion of the signal pulse is used to pump the material, while the mid-IR probe pulses were produced through difference-frequency generation between the signal and idler in an external non-linear crystal ($\rm AgGaS_2$) \cite{petrov2001generation}.  An electrically controlled delay stage is then used to vary the temporal separation between the pump and probe, enabling time-resolved measurement of the transient carrier dynamics.

A schematic of the pulse-generation and the femtosecond pump-probe system are shown in Figure \ref{Figure_2} (a). The pump and probe spectra were measured using Fourier-Transform infrared (FTIR) spectroscopy and the results are shown in Figure \ref{Figure_2}(b). The pump and probe pulses are approximately 100 fs in duration, and have radii ($\rm e^{- 1}$ of the intensity) of $\simeq$100 $\rm \mu m$ and $\simeq$35 $\rm \mu m$, respectively, which ensures that the probe beam samples a spatially uniform region of photo-excitation. The polarization of the probe at the sample position is fixed, however, the pump polarization relative to the probe is controlled by a half wave plate. The sample was mounted on a rotation stage, to control the orientation of the bP flake relative to probe beam polarization. To confirm that the substrate does not contribute to the pump-probe signals, these measurements are performed on bare $\rm CaF_2$ substrate, and no signal was observed in this case.

\begin{figure}[htb]
 \includegraphics[width=3in, angle=0]{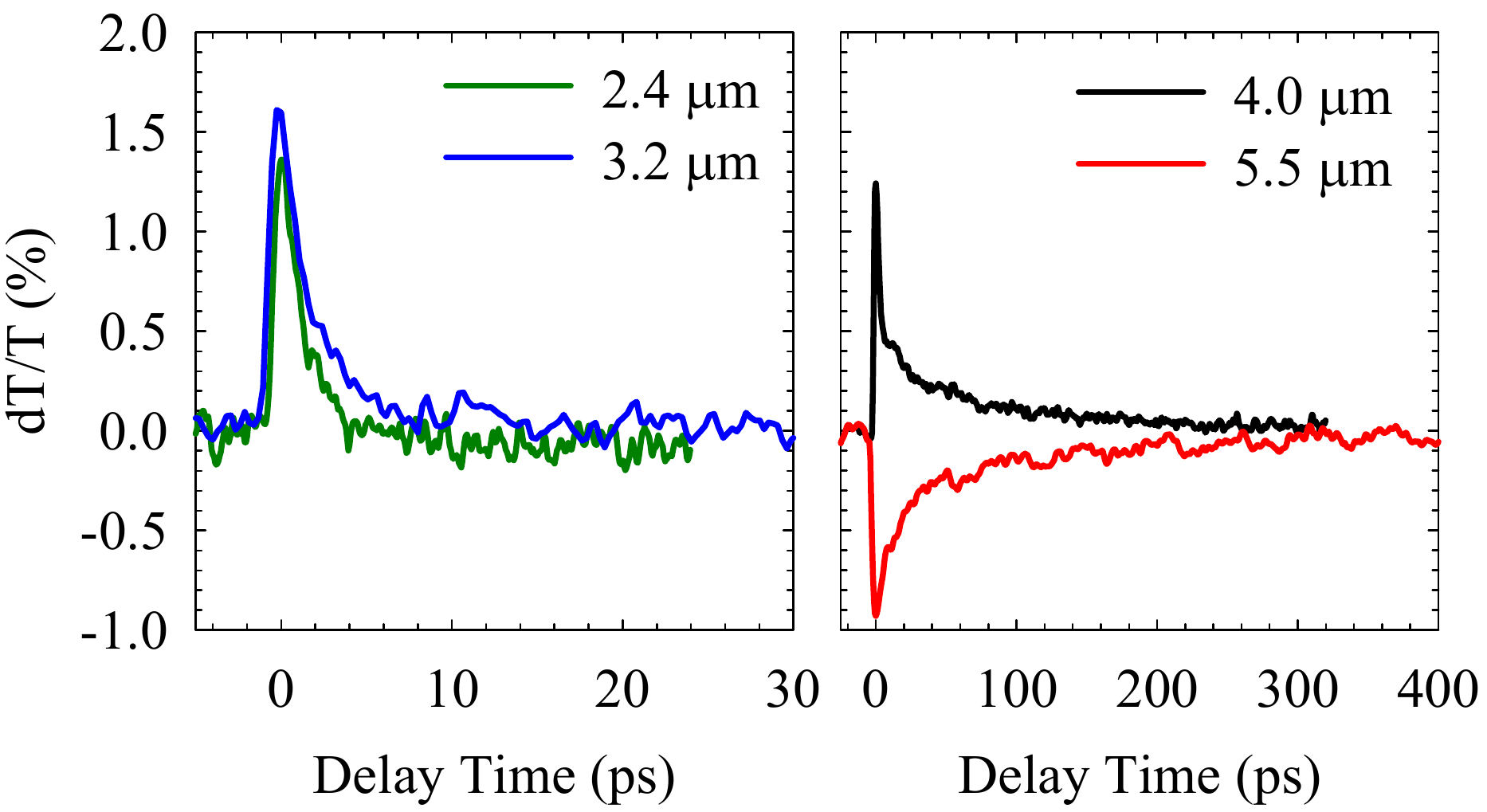}
 \caption{\label{Figure_3} Pump-induced change in transmission as a function of the time delay (at room temperature) for $\sim$ 1.4 $\mu$m pump wavelength and $\sim$ 300 $\rm \mu J /cm^2$ injected pump fluence at 2.4 $\mu$m, 3.2 $\mu$m (on left), 4 $\mu$m and 5.5 $\mu$m (on right) probe wavelengths. The initial development of band filling (t<0) and the full time range recovery (t>0) of the differential transmission as the carriers recombine are plotted.}
\end{figure}

The pump-induced changes in transmission of multi-layer bP at the various probe wavelengths from 2.4 $\mu$m to 5.5 $\mu$m are shown in Figure \ref{Figure_3} (a) and (b). In these measurements, the pump wavelength is set to $\sim$1.4 $\mu$m and the pump intensity is approximately 300 $\rm \mu J /cm^2$. The pump and probe pulses are co-polarized along the arm-chair direction of the multi-layer bP flake in this set of measurements. For probe photon energies larger than the bandgap, Pauli blocking is the dominant nonlinear effect, leading to a pump-induced increase in transmission. Pauli blocking occurs when the final transition states are occupied with photogenerated carriers, thereby blocking the generation of new carriers through direct absorption of the probe pulse.  A fast, transient increase in differential-transmission signal is observed with a measured response time of 1.4 ps and 2 ps for 2.4 $\mu$m and 3.2 $\mu$m probe wavelengths, respectively. These timescales characterize the rate at which pump-induced hot carriers thermalize and relax to the band edge\cite{aytac2014effects}.  When the probe wavelength continues to increase, the differential-transmission signal becomes progressively slower, and when the probe wavelength is extended to 4 $\mu$m, just above the bandgap, the Pauli-blocking transient persists for nearly 90 ps. At this wavelength, the process of band-filling governs the Pauli blocking, as photogenerated carriers accumulate above the band edge before recombining.  Moreover, the free electrons and holes also diminish the transmission because of intraband free-carrier absorption.  In this regime, the overall response is thus governed by competing intra-band and inter-band transitions that increase or decreases the probe transmission, respectively. Hence, the carrier recombination lifetime can be unambiguously estimated only if the probe energy is lower than the bandgap of the material, where inter-band absorption can be conclusively excluded. The red curve in Fig.~\ref{Figure_3}(b) shows the response measured at a probe wavelength of 5.5 $\mu$m, which is strictly negative, and attributed exclusively to (intra-band) free-carrier absorption. According to the Drude mode, the free-carrier absorption coefficient can be expressed as \cite{chandola2005below},
\begin{equation}
\alpha_{\rm FCA}=\frac{q^3 \lambda^2 \Delta n}{4 \pi^2 c^3 \epsilon_0 n_{\rm ind} m^{*2} \mu}
\end{equation}
where $\rm \lambda$ is the wavelength of the incident optical field, $m^*$ is the effective mass of the involved carrier (either electrons or holes), $\mu$ is the mobility of the carrier type, $n_{\rm ind}$ refractive index, and $\Delta n$ is the excess carrier density. The quadratic dependence of FCA on wavelength makes this process most pronounced for longer wavelengths, or in other words, for low energy, sub-bandgap photons. When the photo-excited carriers cool to the band edge, the population of carriers achieves thermal equilibrium at the lattice temperature, but with an excess population of electron-hole pairs in the system.  By probing the system at a wavelength below the bandgap, the recombination lifetime of these photo-excited carriers can be directly measured. By further examining the fluence dependence in this regime, one can reliably measure the carrier recombination lifetimes related to the minority carrier, radiative, and Auger processes. \cite{aytac2017mid}.

To investigate the minority carrier lifetime and radiative recombination processes, we measured the differential-transmission at 5.5 $\rm \mu$m probe wavelength for a range of different pump intensities. The pump and probe pulses are co-polarized and directed along the arm-chair direction of the bP flake.  From the measured pump power ($P_e$), laser repetition rate ($R_r$), and pump spot size radius ($w_e$), the on-axis pump fluence ($F_e$) can be calculated as,
\begin{equation}
F_e=(1-R) \frac{P_e}{\pi w_e^2 R_r}
\end{equation}
where $R$ is the reflection of pump beam from the CaF$_2$  substrate. The initial injected excess carrier density in the bP flake is directly proportional to the pump fluence which can be obtained from,
\begin{equation}
\Delta n=\frac{F_e}{h\upsilon L}\times \left(1-e^{-\alpha_e L}\right),
\end{equation}
where $\rm \alpha_e$ is the absorption coefficient at the pump wavelength, and $L$ is the bP thickness\cite{suess2016mid}. 

\begin{figure*}[htb]
 \includegraphics[width=6in, angle=0]{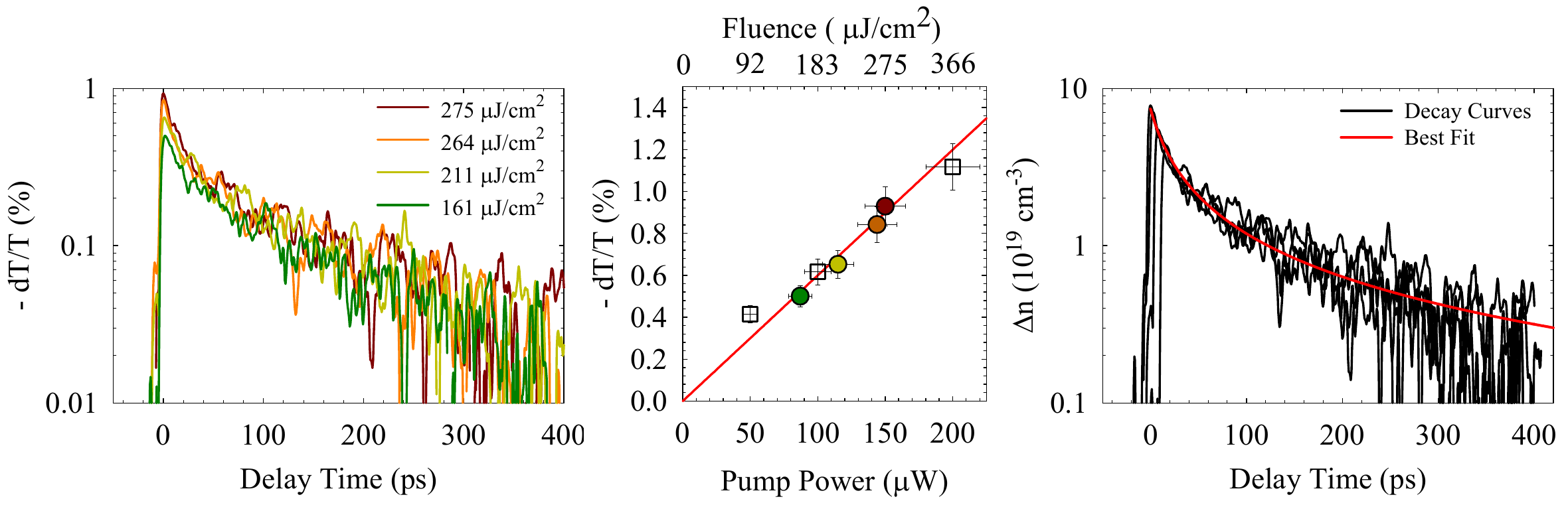}
 \caption{\label{Figure_4} (a) Time-resolved differential-transmission decay curves illustrating carrier recombination in multi-layer bP flake. (b) Peak dT/T, found at zero time delay, as a function of pump power and fluence. (c) Excess carrier density as a function of time is shown with the best fit of recombination model. }
\end{figure*}

Fig.~\ref{Figure_4}(a) shows the time-resolved differential-transmission for injected carriers concentration (fluence) ranging from $\rm \Delta n$ from $\sim$ $\rm 4 \times 10^{19}$  $\rm cm^{-3}$ (161 $\rm \mu J/cm^2$) to $\sim$ $\rm 9 \times 10^{19}$  $\rm cm^{-3}$ (275 $\rm \mu J/cm^2$).  For all fluences, the response is negative, as expected for intra-band absorption.  To better reveal the non-exponential nature of the carrier recombination, the response $-dT/T$ is plotted on semilogarithmic axes.  At high pump fluence, a fast decay is apparent in the first few picoseconds which gradually decrease to a slower recovery time for the lower pump fluences.  The measured peak value of $|dT/T|$ and the corresponding initial photo-generated excess carrier density are linearly proportional for incident pump powers up to 300 $\rm \mu J/cm^2$ as shown in Figure \ref{Figure_4} (b). Using this proportionality, we convert the measured the time-resolved differential-transmission data to the time-resolved excess carrier density data, $\Delta n(t)$, which is shown in Figure \ref{Figure_4}(c). Additionally, the four individual decay curves are merged into one by appropriately shifting the zero-time position of the curves.  This suggests that the carrier recombination can be described by a time-invariant first-order nonlinear differential equation, in which the subsequent transient is determined exclusively by one initial condition: the carrier concentration.

At the lowest photo-generated carrier densities, the response approaches an exponential decay, indicating that for low concentration the recombination rate is in proportional to $\Delta n$. At the largest photo-generated carrier densities, a dramatically faster decay is evident in the first few picoseconds that has been previously attributed to radiative recombination in bP \cite{suess2016mid,bhaskar2016radiatively}. The total recombination rate can be expressed as,
\begin{equation}
\frac{\partial \Delta n}{\partial t}= -A \Delta n - B \Delta n^2.
\end{equation}
Here, $A$ is the non-radiative and $B$ is the radiative recombination coefficient. This nonlinear equation can be directly integrated to obtain the solution,
\begin{equation}
\Delta n(t)=\left( \Delta n(t=0)^{-1} + \frac{B}{A} (e^{A t}-1) \right)^{-1}
\end{equation}
The initial excess carrier density injected into the system by the pump pulse is denoted $\Delta n(t=0)$. The minority carrier lifetime $A^{-1}$ of 1.34 ns and the radiative recombination coefficient $B$ of 5.9 $\rm \times 10^{-10} cm^3/s$ are determined from the best fit to the data shown, shown by the red curve in Fig.~\ref{Figure_4}(c). These results are consistent with previously reported values obtained from time-resolved photo-conductivity \cite{suess2016mid} and time-resolved micro-wave conductivity \cite{bhaskar2016radiatively} measurements, which report minority carrier lifetime of approximately 1 ns and the radiative recombination coefficient of approximately 2 $\rm \times 10^{-10} cm^3/s$.   The higher radiative recombination rate observed in our experiments may be attributed to differences in the thickness, geometry, and purity of the bP flake considered here. These numbers are comparable in magnitude to other narrow bandgap materials such as InAs, InSb, and GaSb \cite{aytac2017mid,Donetsky.2010,levinshtein1996handbook}.

In conclusion, the carrier lifetime of multi-layer bP is investigated at room temperature, using two-color pump-probe measurements. By probing the pump-induced change in transmission through the bP flake at photon energies above and below the bandgap, we can assess the different roles of inter-band and intra-band processes. This spectroscopically comprehensive measurement allows for clear determination of carrier dynamics in bP at the different levels of the electronic band structure. The actual carrier lifetime is measured at a wavelength of 5.5 $\mu$m with photon energies well below the bandgap. The use of various photon energies above the bandgap allowed estimating the carrier redistribution time via thermalization and relaxation to the bottom of conduction band. An excess carrier density dependent recombination rate function used to estimate minority carrier and radiative recombination rates 1.3 $\rm ns^{-1}$  and  5.9 $\rm \times 10^{-10} cm^3/s$. The high radiative coefficient and short minority carrier lifetime (high internal quantum efficiency) suggest that bP light-emitting devices could be promising for mid-IR photonics.

%\begin{acknowledgments}

%\end{acknowledgments}

%\begin{thebibliography}{20}%

%\bibliography{Refs_522018}

\end{document}